\documentclass[letterpaper]{article} 
\usepackage{aaai24}  
\usepackage{times}  
\usepackage{helvet}  
\usepackage{courier}  
\usepackage[hyphens]{url}  
\usepackage{graphicx} 
\usepackage{natbib}  
\usepackage{caption} 
\usepackage{amsmath,amsfonts,amsthm, amssymb}
\newtheorem{theorem}{Theorem}

\newtheorem{definition}{Definition}
\newtheorem{proposition}{Proposition}
\usepackage{algorithm}
\usepackage{algorithmic}
\usepackage{multirow}

\usepackage{newfloat}
\usepackage{listings}
\usepackage[inkscapelatex=false]{svg}
\DeclareCaptionStyle{ruled}{labelfont=normalfont,labelsep=colon,strut=off} 
\floatstyle{ruled}
\newfloat{listing}{tb}{lst}{}
\floatname{listing}{Listing}
\title{Robust Beamforming for Downlink Multi-Cell Systems: \\ A Bilevel Optimization Perspective}
\author{
    Xingdi Chen\textsuperscript{\rm 1},
    Yu Xiong\textsuperscript{\rm 1},
    Kai Yang\textsuperscript{\rm 1,\rm 2,\rm 3}\thanks{Corresponding author.}
}
\affiliations{
    \textsuperscript{\rm 1}Department of Computer Science and Technology, Tongji University, China\\
    \textsuperscript{\rm 2}Key Laboratory of Embedded System and Service Computing Ministry of Education at Tongji University\\
    \textsuperscript{\rm 3}Shanghai Research Institute for Intelligent Autonomous Systems

    xingdichen@tongji.edu.cn, septerxy@tongji.edu.cn, kaiyang@tongji.edu.cn
}

\usepackage{bibentry}

\begin{document}

\maketitle

\begin{abstract}
Utilization of inter-base station cooperation for information processing has shown great potential in enhancing the overall quality of communication services (QoS) in wireless communication networks. Nevertheless, such cooperations require the knowledge of channel state information (CSI) at base stations (BSs), which is assumed to be perfectly known. However, CSI errors are inevitable in practice which necessitates beamforming techniques that can achieve robust performance in the presence of channel estimation errors. Existing approaches relax the robust beamforming design problems into semidefinite programming (SDP), which can only achieve a solution that is far from being optimal. To this end, this paper views robust beamforming design problems from a bilevel optimization perspective. In particular, we focus on maximizing the worst-case weighted sum-rate (WSR) in the downlink multi-cell multi-user multiple-input single-output (MISO) system considering bounded CSI errors. We first reformulate this problem into a bilevel optimization problem and then develop an efficient algorithm based on the cutting plane method. A distributed optimization algorithm has also been developed to facilitate the parallel processing in practical settings. Numerical results are provided to confirm the effectiveness of the proposed algorithm in terms of performance and complexity, particularly in the presence of CSI uncertainties.

\end{abstract}

\section{Introduction}
In multi-cell multiuser wireless communication networks, users, especially cell-edge users, may have low-rate data service as a consequence of suffering from both intra-cell interference and inter-cell interference. Beamforming, one of the most promising multi-antenna techniques, harnesses the spatial dimension to effectively alleviate interference in downlink transmissions. This technique is extensively employed in the design of mobile communication systems. However, such precoding technique is contingent upon hardware complexity, often rendering it impractical for mobile terminals characterized by limited computational power and storage capacity. Consequently, multiple transmit antennas are only deployed at the base station (BS) where the issue of computing power is less problematic, while mobile units are equipped with a small number of antennas. This configuration is commonly referred to as a multiple-input and multiple-output (MIMO) system. Beamforming techniques inherently necessitate channel state information (CSI) at each BS to enable precoded transmissions.

Various quality of communication services (QoS) metrics are used in wireless networks, including maximum allowable mean square errors, minimum tolerated signal-to-interference-plus-noise ratios, and weighted sum rates. One widely used QoS metric is the weighted sum-rate (WSR), making it a fundamental and extensively studied problem to design linear beamformers that maximize WSR under total power constraints. Since the WSR maximization problem is nonconvex and NP-hard even in the single-antenna case \cite{4453890}, it is challenging to achieve the global optimal solution and suboptimal solutions are of great interest. Assuming perfect CSI, \cite{6832894} points out that the optimal beamforming vectors for the WSR maximization problem in single-cell downlink transmission have a simple structure. Several algorithms have been proposed for single-cell downlink transmission \cite{6170850, 6119233}. However, extending the WSR maximization problem to multi-cell downlink transmission poses greater challenges. An iterative beamformer design based on iterative second-order cone programming (SOCP) approximation is introduced in \cite{6327333}. Additionally, there are also several distributed methods for multi-cell systems \cite{6153325, 6331563, 5756489, 6104172}. Specifically, \cite{6153325} presents a fully distributed beamforming technique relying on the high signal-to-interference-plus-noise ratio (SINR) assumption. This technique exclusively utilizes local CSI without requiring additional information exchange. \cite{6331563} splits the nonconvex problem into a master problem that is addressed by a novel sequential convex approximation, and multiple subproblems that can be solved by BSs in a fully asynchronous manner through the primal decomposition technique. Both methods in \cite{6153325} and \cite{6331563} are restricted to multiple-input single-output (MISO) systems. The algorithms proposed in \cite{5756489} and \cite{6104172} are both based on iterative minimization of mean-square error (MMSE) and can tackle the WSR maximization problem in MIMO environments. 

Regrettably, achieving perfect CSI at transmitters is unattainable due to estimation and quantization errors. The limited amount of feedback bits available for feeding back CSI gives rise to quantization errors, which are the dominant source of the uncertainty in CSI \cite{5590310}. Additionally, the provision of up-to-date CSI at transmitters also remains questionable. Therefore, the consideration of imperfect CSI, known to be detrimental to the performance of those methods assuming perfect CSI, becomes crucial. This practical constraint has led to the emergence of robust beamforming techniques, aiming to guarantee the worst-case performance of the network under CSI imperfections.

Two main approaches exist for modeling the uncertainty region of CSI errors. The first approach employs the probabilistic model, treating errors as a random variable with some known distribution \cite{1605472, 1665009, 4567684, 4586299, 7555358}. Most of these works aim at optimizing a utility function by averaging over the entire uncertainty region. In this paper, we adhere to the second approach, wherein the uncertainty region of CSI perturbations is confined within given bounded uncertainty sets \cite{4663943, 5590310, 6156468, 6824195, 9110587}. Making no assumption on the distribution of CSI errors, this approach matches well with the quantization errors. Moreover, this method also works for unbounded errors as long as the system outage probability is controlled. 

Incorporating CSI imperfections leads to the formulation of robust optimization problems. \cite{4663943} minimizes the transmit power under a predetermined set of QoS constraints for users. In the presence of bounded CSI errors, the constraints are infinite due to the fact that the QoS requirements must be supported for an infinite number of possible channels within the uncertainty regions. To address this issue, authors employ  semidefinite programming (SDP) along with a lemma to convert an infinite set of constraints into a finite set of constraints, making the problem computationally tractable. Both \cite{6156468} and \cite{6824195} adopt techniques like semidefinite relaxation (SDR) and the S-Lemma to reformulate the original optimization problem into a numerically tractable one. They then propose distributed algorithms based on ADMM and primal decomposition approaches, respectively. For the problem of maximizing the worst-case WSR of the network, \cite{5590310} provides a lower-bound solution by introducing an additional function and then transforming the problem into the weighted sum of the worst-case mean-square error minimization problem that can be solved by SDP. However, as the system scales up, for example, with an increase in the number of cells and the number of transmitting antennas on BSs, these algorithms become impractical due to the time-consuming nature of SDP.

Bilevel optimization dates back to the literature \cite{hicks1935marktform}. Recently, bilevel optimization has gained significant attention and is widely applied in various machine learning applications including wireless communication \cite{9682542}, hyperparameter optimization\cite{liu2021value, franceschi2018bilevel}, meta learning \cite{ji2020convergence, ji2021bilevel} and neural architecture search \cite{liu2018darts, xue2021rethinking}. Bilevel optimization is an optimization problem where a subset of variables is constrained to be optimal for another given optimization problem. Mathematically, a general bilevel optimization takes the following formulation, 
\begin{equation}
\begin{array}{ll}
\underset{\boldsymbol{x}}{\min}  &F(\boldsymbol{x}, \boldsymbol{y}) \\
\text { s.t. }  & G(\boldsymbol{x}, \boldsymbol{y}) \leq 0 \\
 &\boldsymbol{y} \in \underset{\boldsymbol{y}^{\prime} \in \mathcal{Y}} {\arg \min} \left\{f\left(\boldsymbol{x}, \boldsymbol{y}^{\prime}\right) \mid g\left(\boldsymbol{x}, \boldsymbol{y}^{\prime}\right) \leq 0\right\}, \\
\end{array}
\end{equation}
where $F$ and $f$ denote the upper-level and lower-level objective functions, respectively. $\boldsymbol{x} \in \mathbb{R}^n$ is the upper-level variable and $\boldsymbol{y} \in \mathbb{R}^m$ is the lower-level variable. We refer to $G$ and $g$ as the upper-level constraint and the lower-level constraint, respectively. In most existing bilevel optimization works in machine learning tasks, both upper-level constraint and lower-level constraint are not considered due to the characteristics of tasks\cite{liu2021value, franceschi2018bilevel, ji2020convergence, ji2021bilevel, liu2018darts, xue2021rethinking}. To the best of our knowledge, bilevel optimization has not been applied to robust beamforming designs. This suggests a promising avenue for exploration in adapting bilevel optimization methods to address robust beamforming problems.

In this paper, we consider multi-cell multiuser MISO wireless networks \cite{5590310}. The main problem of interest is the beamforming optimization with the goal of maximizing the WSR within per BS power constraints in the presence of CSI imperfections. To ensure the worst-case performance, we assume CSI errors are bounded and resort to robust optimization \cite{4509766, yang2014distributed}. For obtaining such beamformers, we begin by transforming the original robust optimization problem into a bilevel optimization problem, encompassing both upper-level and lower-level constraints. Secondly, we develop a \textbf{B}i\textbf{L}evel based \textbf{R}obust \textbf{B}eam\textbf{F}orming (BLRBF) algorithm similar to the centralized method proposed in Appendix A of the reference \cite{jiao2022asynchronous}. To be specific, we treat the lower-level optimization problem as a constraint to the upper-level optimization problem and utilize cutting planes to approximate this constraint. Subsequently, inspired by the work \cite{6600797}, we extend the BLRBF method to the asynchronous distributed implementation in order to faster approximate the feasible region. This distributed algorithm is referred to as \textbf{B}i\textbf{L}evel based \textbf{A}synchronous \textbf{D}istributed \textbf{R}obust \textbf{B}eam\textbf{F}orming (BLADRBF). Notably, our algorithm can be readily extended to MIMO systems. We prove that both BLRBF and BLADRBF are guaranteed to converge.

\textbf{Contributions}. Our main contributions are summarized as follows:
\begin{itemize}
\item{We are the first to propose viewing robust beamforming design problems from a bilevel optimization perspective. Unlike conventional methods that rely on SDP, which are computationally expensive and can only achieve a solution that is far from being optimal. The fresh perspective provides new insights into solving such problems and offers a promising alternative with the potential for improved performance.}
\item{To illustrate the application of bilevel optimization, we present a novel bilevel based formulation and develop a cutting plane based algorithm called BLRBF. This approach efficiently handles the challenging task of maximizing worst-case weighted sum-rates. }
\item{We also propose an asynchronous distributed algorithm (BLADRBF) to facilitate the parallel processing in practical settings. The asynchronism gives the algorithm a high robustness against failures in the communication. Importantly, both algorithms are mathematically proven to converge.}
\end{itemize}


\section{System Model}
In this section, we consider a multi-cell MISO downlink system with $M$ cells each equipped with one BS with $N$ antennas that serves single-antenna $K$ users. The BS of the $m$th cell and the $k$th user in the $m$th cell are denoted by $\text{B}_m$ and $\text{U}_{k_m}$, respectively. The transmitted signal from $\text{B}_m$ is given by
\begin{equation}
    \boldsymbol{x}_m = \sum_{k=1}^{K} \boldsymbol{v}_{k_m} s_{k_m},
\end{equation}
where $\boldsymbol{v}_{k_m} \in \mathbb{C}^{N \times 1}$ represents the beamformer that the $m$th BS uses to transmit the signal $s_{k_m} \sim \mathcal{CN}(0, 1)$ to user $\text{U}_{k_m}$ and we assume $\mathbb{E}[|s_{k_m}|^2]=1$. Then, the received signal at $\text{U}_{k_m}$ is given by
\begin{equation}
\begin{aligned}
    y_{k_m} = 
    \underbrace{ \boldsymbol{h}_{k_m m} \boldsymbol{v}_{k_m}s_{k_m}}_{\text{the desired signal}}
    &+\underbrace{\sum_{l \neq k} \boldsymbol{h}_{k_m m} \boldsymbol{v}_{l_m}s_{l_m}}_{\text{intra-cell interference}}  \\
    &+\underbrace{\sum_{n \neq m} \sum_{l} \boldsymbol{h}_{k_m n} \boldsymbol{v}_{l_n}s_{l_n} + n_{k_m}}_{\text{inter-cell interference plus noise}},
\end{aligned}
\end{equation}
where $\boldsymbol{h}_{k_m n} \in \mathbb{C}^{1 \times N}$ represents the channel from  $\text{BS}_{n}$ to $\text{U}_{k_m}$ and $n_{k_m} \sim \mathcal{CN}(0, \sigma_{k_m}^2)$ denotes the additive complex white Gaussian noise of $\text{U}_{k_m}$. Accordingly, the  SINR of $\text{U}_{k_m}$ can be written as
\begin{equation}
\begin{aligned}
&\operatorname{SINR}_{k_m}  ( \left \{ \boldsymbol{v}_{1_m},\cdots,\boldsymbol{v}_{K_m} \right \}_{m = 1}^{M}   ) \\
&=  \frac{\left|\boldsymbol{h}_{k_m m} \boldsymbol{v}_{k_m}\right|^{2}}{\sum_{l \neq k}\left| \boldsymbol{h}_{k_m m} \boldsymbol{v}_{l_m} \right|^{2}+\sum_{n \neq m} \sum_{l}\left| \boldsymbol{h}_{k_m n} \boldsymbol{v}_{l_n}\right|^{2}+\sigma_{k_m}^2}.
\end{aligned}
\end{equation}
It is assumed that the $\text{B}_m$ knows only erroneous channel estimates \{$\widetilde{\boldsymbol{h}}_{k_m n}$\}, i.e.,
\begin{equation}
\begin{array}{c}
    \boldsymbol{h}_{k_m n} = \widetilde{\boldsymbol{h}}_{k_m n} + \hat{\boldsymbol{\Delta}}_{k_m n}, \\
    \forall m,n \in \{1,...,M\}, \quad \text{and} \quad  \forall k \in \{ 1,...,K \},
\end{array}
\end{equation}
where $\hat{\boldsymbol{\Delta}}_{k_m n}$ is the channel estimation errors, which are unknown to BSs. Furthermore, the BSs are supposed to know the structure of the uncertainty regions, which in this paper, are bounded and defined as origin-centered hyper-spherical region of radius $\epsilon_{k_m n}$, i.e., $\| \hat{\boldsymbol{\Delta}}_{k_m n} \|_{2} \leq \epsilon_{k_m n}$. 

For notational simplicity, we denote the beamformer of $\text{B}_m$ by $\hat{\boldsymbol{V}}_m=[\boldsymbol{v}_{1_m},\boldsymbol{v}_{2_m},\ldots,\boldsymbol{v}_{K_m}] \in \mathbb{C}^{N \times K}$. Then, the problem of interest is to find the transmit beamformers \{$\hat{\boldsymbol{V}}_m$\} such that the worst-case WSR of the network is maximized, while the power of each BS is constrained. Mathematically, this problem can be formulated as
\begin{equation}\label{problem}
\begin{array}{cl}
    \underset{\left \{ \hat{\boldsymbol{V}}_{m} \right \} }\max \underset{\left \{ \hat{\boldsymbol{\Delta}}_{k_m n} \right \}}\min
    & \sum_{m=1}^{M} \sum_{k=1}^{K} \alpha_{k_m} \log \left(1+\operatorname{SINR}_{k_m}\right) \\
    \operatorname{s.t.}
    & \left\| \hat{\boldsymbol{V}}_{m}\right\|_{F}^{2} \leq P_{m} \quad \forall m ,\\
\end{array}
\end{equation}
where $ \alpha_{k_m}$ is the positive weighting factor corresponding to the rate of $\text{U}_{k_m}$ and $P_m$ is the power budget of the BS $\text{B}_m$. The objective function of the problem (\ref{problem}) is known as the WSR utility function. This problem focuses solely on maximizing the throughput of the network, disregarding the minimum rate requirements of individual users. This utility function is proved to be highly suitable for scenarios where users can tolerate delays due to the fact that certain users may not receive any resources during specific scheduling frames for the benefit of the network throughput in the extreme case \cite{1665009}. The incorporation of weighting factors enables us to assign different priorities to individual users, thereby allowing us to cater to diverse user needs. Additionally, these weights can be dynamically adjusted over time to ensure long term fairness.

The WSR utility function is nonconvex and involves CSI and beamformers of all BSs. Thus, this optimization problem is quite complicated.

\section{BLRBF and BLADRBF Methods}
In this section, we propose a \textbf{B}i\textbf{L}evel based \textbf{A}synchronous \textbf{D}istributed \textbf{R}obust \textbf{B}eam\textbf{F}orming (BLADRBF) method to solve the problem (\ref{problem}) in an asynchronous distributed manner. We first transform the problem (\ref{problem}) into a bilevel optimization problem and then introduce the centralized method, i.e., BLRBF, similar to the method CPBO proposed in Appendix A of the reference \cite{jiao2022asynchronous}. CPBO deals with problems where there are no constraints at both the upper and lower levels. Unlike them, our problem has both convex upper-level constraints and convex lower-level constraints. Finally, we extend this centralized method to the asynchronous distributed implementation, i.e., BLADRBF.

\subsection{BLRBF: Bilevel Based Robust Beamforming}

For notational simplicity, we split $\{ \hat{\boldsymbol{V}}_{m} \}$ into real and imaginary components and then arrange these components into a single vector, i.e., $\boldsymbol{V} \triangleq [\text{Vec}(Re(\boldsymbol{\hat{V}})),\text{Vec}(Im(\boldsymbol{\hat{V}}))] \in \mathbb{R}^{2MNK \times 1}$. And $\boldsymbol{\Delta} \in \mathbb{R}^{2 M^2 N K \times 1}$ is defined in the same way. 

First, we define 
\begin{equation}
    f( \boldsymbol{V}, \boldsymbol{\Delta}) = \sum_{m=1}^{M} \sum_{k=1}^{K} \alpha_{k_m} \log \left(1+\operatorname{SINR}_{k_m}\right).
\end{equation}
Then, from the perspective of bilevel optimization, problem (\ref{problem}) can be written as
\begin{equation}\label{problem_bilevel}
\begin{array}{ll}
    \underset{\boldsymbol{V}}{\min}  &-f(\boldsymbol{V},  \boldsymbol{\Delta} ) \\
    \text { s.t. }  & \| \hat{\boldsymbol{V}}_{m}\|_{F}^{2} \leq P_{m}, \quad \forall m \\
     &\boldsymbol{\Delta} = \underset{\boldsymbol{\Delta}^{\prime} } {\arg \min} f\left( \boldsymbol{V} ,\boldsymbol{\Delta}^{\prime} \right)\\
     &\qquad \| \hat{\boldsymbol{\Delta}}_{k_m n}^{\prime} \|_{2} \leq \epsilon_{k_m n}, \quad \forall m,n,k,  \\
\end{array}
\end{equation}
where specially, the upper and lower objective functions differ only by a negative sign. 

Now, we begin the process of solving this bilevel optimization problem. By defining $\phi(\boldsymbol{V}) = \arg \min_{ \boldsymbol{\Delta}^{\prime} } \left\{f\left( \boldsymbol{V} , \boldsymbol{\Delta}^{\prime}\right) \mid  \| \hat{\boldsymbol{\Delta}}_{k_m n}^{\prime} \|_{2} \leq \epsilon_{k_m n}, \forall m,n,k\right\}$ and $g(\boldsymbol{V} ,\boldsymbol{\Delta} )=\| \boldsymbol{\Delta} -\phi( \boldsymbol{V})\|_{2}^{2}$, we can reformulate the problem (\ref{problem_bilevel}) as a single-level problem
\begin{equation}\label{single optimization}
\begin{array}{cl}
    \underset{ \boldsymbol{V},  \boldsymbol{\Delta}}{\min}  &-f( \boldsymbol{V} , \boldsymbol{\Delta} ) \\
    \text { s.t. }  & \| \hat{\boldsymbol{V}}_{m}\|_{F}^{2} \leq P_{m}, \quad \forall m \\
     &g(\boldsymbol{V} ,\boldsymbol{\Delta} ) =0.
\end{array}
\end{equation}

In order to get an estimate of $\phi(\boldsymbol{V})$, we first transform the inequality lower-level constraints into equality constraints by introducing slack variables $ \left\{s_{k_m n} \right\}$, and then turn to augmented Lagrangian method. In specific, the lower-level problem can be given by
\begin{equation}\label{lower problem}
\begin{array}{cl}
     \underset{ \boldsymbol{\Delta}^{\prime} ,  \left\{s_{k_m n} \right\}  }{\arg \min} & f(\boldsymbol{V} , \boldsymbol{\Delta}^{\prime}  )\\
     \text{s.t.} &\| \hat{\boldsymbol{\Delta}}_{k_m n}^{\prime} \|_{2} + s_{k_m n}^{2} =\epsilon_{k_m n}, \quad \forall m,n,k.
\end{array}
\end{equation}
Considering the first-order Taylor approximation of $f(\boldsymbol{V} , \boldsymbol{\Delta}^{\prime}  )$ with respect to $\boldsymbol{V}$, i.e., for a given point $ \widetilde{\boldsymbol{V}}$, $\tilde{f} (\boldsymbol{V},  \boldsymbol{\Delta}^{\prime})=f(\widetilde{\boldsymbol{V}} , \boldsymbol{\Delta}^{\prime} )+\nabla_{ \boldsymbol{V}}f(\widetilde{\boldsymbol{V}} , \boldsymbol{\Delta}^{\prime} )^T( \boldsymbol{V} - \widetilde{\boldsymbol{V}})$, the augmented Lagrangian function of the lower-level optimization problem (\ref{lower problem}) can be written as
\begin{equation}
\begin{array}{l}
f_{\text{ALM}}( \boldsymbol{V}, \boldsymbol{\Delta}^{\prime},\left\{s_{k_m n} \right\}, \left\{\mu_{k_m n} \right\})=\\
 \tilde{f}(\boldsymbol{V}, \boldsymbol{\Delta}^{\prime})\\
+\sum_{k,m,n}\mu_{k_m n}(\| \hat{\boldsymbol{\Delta}}_{k_m n}^{\prime} \|_{2} + s_{k_m n}^{2} -\epsilon_{k_m n})\\
+\sum_{k,m,n}\frac{\rho}{2}(\| \hat{\boldsymbol{\Delta}}_{k_m n}^{\prime} \|_{2} + s_{k_m n}^{2} -\epsilon_{k_m n})^{2},
\end{array}
\end{equation}
where $\left\{\mu_{k_m n} \right\}$ are Lagrange multipliers, and $\rho > 0$ is the penalty parameter. Therefore, based on augmented Lagrangian method, we have
\begin{equation}\label{ALM}
\begin{array}{rl}
\boldsymbol{\Delta}^{\prime}_{k+1} &=  \boldsymbol{\Delta}^{\prime}_{k}-\eta _{\boldsymbol{\Delta}^{\prime}}\nabla _{\boldsymbol{\Delta}^{\prime}}f_{\text{ALM}k},\\
\left \{ s_{k_m n} \right \}_{k+1} &= \left \{ s_{k_m n} \right \}_{k}-\eta _{\left \{ s_{k_m n} \right \}}\nabla _{\left \{ s_{k_m n} \right \}}f_{\text{ALM}k},\\
\left \{ \mu_{k_m n} \right \}_{k+1} &= \left \{ \mu_{k_m n} \right \}_{k}+\eta _{\left \{ \mu_{k_m n} \right \}}\nabla _{\left \{ \mu_{k_m n} \right \}}f_{\text{ALM}k},\\
\end{array}
\end{equation}
where $f_{\text{ALM}k}=f_{\text{ALM}}(\boldsymbol{V}, \boldsymbol{\Delta}^{\prime}_{k},\left\{s_{k_m n} \right\}_{k}, \left\{\mu_{k_m n} \right\}_{k})$, and $\eta_{\boldsymbol{\Delta}^{\prime}}$, $\eta _{\left \{ s_{k_m n} \right \}}$, $\eta _{\left \{ \mu_{k_m n} \right \}}$ are step-sizes. If we repeat procedure (\ref{ALM}) $K$ times, $\phi(\boldsymbol{V})$ can be approximated by
\begin{equation}\label{phiV}
\phi(\boldsymbol{V})= \boldsymbol{\Delta}^{\prime}_{0}-\sum_{k=0}^{K-1}\eta _{\boldsymbol{\Delta}^{\prime}}\nabla _{\boldsymbol{\Delta}^{\prime}}f_{\text{ALM}k}.
\end{equation}

Let us then consider the relaxed problem of (\ref{single optimization})
\begin{equation}\label{relaxed single optimization}
\begin{array}{cl}
    \underset{ \boldsymbol{V},  \boldsymbol{\Delta}}{\min}  &-f( \boldsymbol{V} , \boldsymbol{\Delta} ) \\
    \text { s.t. }  & \| \hat{\boldsymbol{V}}_{m}\|_{F}^{2} \leq P_{m}, \quad \forall m  \\
     &g(\boldsymbol{V} ,\boldsymbol{\Delta} ) \leq \varepsilon,
\end{array}
\end{equation}
where $\varepsilon > 0$ is a very small constant. The feasible region of this problem is defined as $\mathcal{S}$.


\begin{theorem} \label{Theorem_convex}
The function $g(\boldsymbol{V}, \boldsymbol{\Delta})$ is convex with respect to $(\boldsymbol{V}, \boldsymbol{\Delta})$ and the feasible region $\mathcal{S}$ is convex when $K=1$. 
\end{theorem}
\noindent The proof of Theorem \ref{Theorem_convex} is presented in Appendix A.

We use a set of cutting planes to approximate the convex feasible region. These cutting planes forms a polytope $\mathcal{D}^{[t]}$ at $t$-th iteration, which can be given by
\begin{equation}
    \mathcal{D}^{[t]}=\left \{ \boldsymbol{a}_i^T \boldsymbol{V} + \boldsymbol{b}_i^T \boldsymbol{\Delta} + \kappa_i \leq0,i=1,...,|\mathcal{D}^{[t]}|\right \} ,
\end{equation}
where $\boldsymbol{a}_i \in \mathbb{R}^{2 M  N  K \times 1}$, $\boldsymbol{b}_i \in \mathbb{R}^{2  M^2 N  K \times 1} $ and $\kappa_i \in  \mathbb{R}^{1}$ are parameters of $i$-th cutting plane and $|\mathcal{D}^{[t]}|$ denotes the number of cutting planes in $\mathcal{D}^{[t]}$. Thus, the problem (\ref{relaxed single optimization}) can be expressed as follows
\begin{equation}\label{CP}
\begin{array}{cl}
    \underset{ \boldsymbol{V},  \boldsymbol{\Delta}}{\min}  &-f( \boldsymbol{V} , \boldsymbol{\Delta} ) \\
    \text { s.t. }  & \boldsymbol{a}_i^T \boldsymbol{V} + \boldsymbol{b}_i^T \boldsymbol{\Delta} + \kappa_i \leq 0, \forall i=1,...,|\mathcal{D}^{[t]}|.
\end{array}
\end{equation}
The Lagrangian function of problem (\ref{CP}) can be written as
\begin{small}
\begin{equation}
L\left(\boldsymbol{V}, \boldsymbol{\Delta},\left\{\lambda_{i}\right\}\right)=
-f(\boldsymbol{V}, \boldsymbol{\Delta})+\sum_{i=1}^{\left|\mathcal{D}^{t}\right|} \lambda_{i}\left(\boldsymbol{a}_{i}^{\top} \boldsymbol{V}+\boldsymbol{b}_{i}^{\top} \boldsymbol{\Delta}+\kappa_{i}\right),
\end{equation}
\end{small}
where $\left\{ \lambda_i \right\}$ are Lagrange multipliers associated with inequality constraints. Thus, the algorithm can proceed in the $t$-th iteration as follows
\begin{equation}\label{update V}
    \boldsymbol{V}^{[t+1]} = \boldsymbol{V}^{[t]}-\eta_{\boldsymbol{V}} \nabla_{\boldsymbol{V}} L\left(\boldsymbol{V}^{[t]}, \boldsymbol{\Delta}^{[t]},\left\{\lambda_{i}^{[t]}\right\}\right),
\end{equation}
\begin{equation}\label{update delta}
    \boldsymbol{\Delta}^{[t+1]} = \boldsymbol{\Delta}^{[t]}-\eta_{\boldsymbol{\Delta}} \nabla_{\boldsymbol{\Delta}} L\left(\boldsymbol{V}^{[t]}, \boldsymbol{\Delta}^{[t]},\left\{\lambda_{i}^{[t]}\right\}\right),
\end{equation}
\begin{equation}\label{update lambda}
    \lambda_{i}^{[t+1]} = \left[ \lambda_{i}^{[t]}+\eta_{\lambda_{i}} \nabla_{\lambda_{i}} L\left(\boldsymbol{V}^{[t]}, \boldsymbol{\Delta}^{[t]},\left\{\lambda_{i}^{[t]}\right\}\right)\right ]_{+}, \forall i,
\end{equation}
where $[\cdot]_{+}$ denotes a projection of a value onto the nonnegative space, and $\eta_{\boldsymbol{V}}$, $\eta_{\boldsymbol{\Delta}}$ and $\eta_{\lambda_{i}}$ are step-sizes. 

\cite{jorge2006numerical} indicates that one of the main challenges in solving constrained optimization problems lies in determining which inequality constraints are active and which are not. By introducing active sets, we can simplify the search for the optimal solution of the problem (\ref{CP}). Because of the complementary slackness, we can decide whether the specific constraint is active through its corresponding dual variable rather than check whether the strict inequality holds \cite{NEURIPS2022_34899013}. To be specific, if $\lambda_{i}>0$, the constraint $\boldsymbol{a}_i^T \boldsymbol{V} + \boldsymbol{b}_i^T \boldsymbol{\Delta} + \kappa_i \leq 0$ is active, and the constraint is inactive if $\lambda_{i}=0$.
To reduce the number of constraints, we can remove inactive constraints during iterations.

Next, we will introduce how to update cutting planes. The cutting planes will be updated every $k_{pre}$ iteration by (a) removing inactive cutting planes and (b) adding new cutting planes. Firstly, We remove inactive cutting planes as follows
\begin{equation}\label{drop cp}
     \mathcal{D}^{[t+1]}=\left\{
        \begin{array}{l}
        \text{Drop}\left(\mathcal{D}^{[t]}, c p_{i}\right), \text { if }\lambda_{i}^{[t+1]}=0 \\
        \mathcal{D}^{[t]}, \text { otherwise },
        \end{array}\right.
\end{equation}
where $cp_i$ represents the $i$-th cutting plane in $\mathcal{D}^{[t]}$ and $\text{Drop}\left(\mathcal{D}^{[t]}, cp_{i}\right) $ means removing the $i$-th cutting plane $cp_i$ from $\mathcal{D}^{[t]}$. As $\mathcal{D}^{[t]}$ is updated, corresponding Lagrange multipliers are supposed to be removed. 

Secondly, we introduce the addition of new cutting planes. Given a query point $(\boldsymbol{V}^{[t+1]}, \boldsymbol{\Delta}^{[t+1]})$, we check whether this point satisfies the constraints $\| \hat{\boldsymbol{V}}_{m} \|_{F}^{2}\leq P_m, \forall m$. If not, we are supposed to generate a new cutting plane to separate the point $(\boldsymbol{V}^{[t+1]}, \boldsymbol{\Delta}^{[t+1]})$ from $\mathcal{S}$. Since the function $\| \hat{\boldsymbol{V}}_{m} \|_{F}^{2}- P_m$ is convex, we can generate valid cutting planes for all $m$ according to \cite{boyd2007localization} as follows
\begin{equation}\label{add cp1}
    \begin{split}
        &\| \hat{\boldsymbol{V}}_{m}^{[t+1]} \|_{F}^{2} - P_{m}+ \\
    &\left[
        \begin{array}{c}
            \frac{\partial \left( \| \hat{\boldsymbol{V}}_{m}^{[t+1]} \|_{F}^{2} - P_{m} \right)}{\partial \boldsymbol{V}} \\
            \boldsymbol{0}
        \end{array}
        \right]^{\top}
    \left(
    \left[
        \begin{array}{l}
            \boldsymbol{V} \\
            \boldsymbol{\Delta}
        \end{array}
        \right]-\left[
        \begin{array}{l}
            \boldsymbol{\boldsymbol{V}}^{[t+1]} \\
            \boldsymbol{\Delta}^{[t+1]}
        \end{array}
        \right]\right) \leq 0,
    \end{split}
\end{equation}
which is denoted by $cp^{[t+1]}_{new,Vm}$. We also need to check whether the point $(\boldsymbol{V}^{[t+1]}, \boldsymbol{\Delta}^{[t+1]})$ is a feasible solution of the problem (\ref{relaxed single optimization}). If not, that is $g(\boldsymbol{V}^{[t+1]}, \boldsymbol{\Delta}^{[t+1]})>\varepsilon$, a valid cutting plane is generated as follows
\begin{equation}\label{add cp2}
    \begin{split}
        &g\left(\boldsymbol{V}^{[t+1]}, \boldsymbol{\Delta}^{[t+1]}\right)+\\
    &\left[
        \begin{array}{c}
            \frac{\partial g\left(\boldsymbol{V}^{[t+1]}, \boldsymbol{\Delta}^{[t+1]}\right)}{\partial \boldsymbol{V}} \\
            \frac{\partial g\left(\boldsymbol{V}^{[t+1]}, \boldsymbol{\Delta}^{[t+1]}\right)}{\partial \boldsymbol{\Delta}}
        \end{array}
        \right]^{\top}
    \left(
    \left[
        \begin{array}{l}
            \boldsymbol{V} \\
            \boldsymbol{\Delta}
        \end{array}
        \right]-\left[
        \begin{array}{l}
            \boldsymbol{\boldsymbol{V}}^{[t+1]} \\
            \boldsymbol{\Delta}^{[t+1]}
        \end{array}
        \right]\right) \leq \varepsilon,
    \end{split}
\end{equation}
which is denoted by $cp^{[t+1]}_{new,g}$. Corresponding Lagrange multipliers should also be added. The details of the proposed algorithm are summarized in algorithm \ref{centralized_algo}.

\begin{theorem}\label{Theorem_convergence}
    (Convergence) The optimal objective value of the problem (\ref{CP}) monotonically converges to a constant $\overline{F}$ over the evolution of the algorithm \ref{centralized_algo}.
\end{theorem}
\noindent The proof of Theorem \ref{Theorem_convergence} is presented in Appendix B.

\begin{algorithm}[tb]
\caption{BLRBF: BiLevel based Robust BeamForming.}
\label{centralized_algo}
\textbf{Input}: $\boldsymbol{P}$, $\left \{ \boldsymbol{\widetilde{h}}_{k_m n}, \epsilon_{k_m n} \right \} $. \\
\textbf{Output}: $\boldsymbol{V}, \boldsymbol{\Delta}$.
\begin{algorithmic}[1] 
\STATE Initialization: set $t=0$,  $\boldsymbol{V}^{[0]}$ randomly, $\boldsymbol{\Delta}^{[0]}=0$ and dual variables $\left \{ \lambda_{i}^{[0]} \right \} =0$;
\REPEAT
\STATE  Update variables $\boldsymbol{V}^{t+1}$, $\boldsymbol{\Delta}^{t+1}$, $\left\{ \lambda_{i}^{[t+1]} \right\}$ according to (\ref{update V}), (\ref{update delta}) and (\ref{update lambda});
\IF {$t$ mod $k_{pre}==0$}
    \STATE Remove inactive cutting planes according to (\ref{drop cp}) and corresponding dual variables;
    \STATE Compute an estimate solution $\phi(\boldsymbol{V}^{[t+1]})$ of the lower level problem according to (\ref{phiV});
        \IF {$\| \hat{\boldsymbol{V}}_{m}^{[t+1]} \|_{F}^{2} > P_m, \forall m$}
        \STATE Add new cutting planes according to (\ref{add cp1}) and corresponding dual variables;
        \ENDIF
        \IF {$g(\boldsymbol{V}^{[t+1]}, \boldsymbol{\Delta}^{[t+1]})>\varepsilon$}
        \STATE Add the new cutting plane according to (\ref{add cp2}) and corresponding dual variable;
        \ENDIF
\ENDIF
\STATE $t \leftarrow t+1$;
\UNTIL{convergence}.
\end{algorithmic}
\end{algorithm}

\subsection{BLADRBF}
In this subsection, we extend the centralized algorithm to a asynchronous distributed implementation, which can approximate the feasible region $\mathcal{S}$ of the problem (\ref{relaxed single optimization}) more quickly. Before that, we introduce the concept of communication graph.
\begin{definition}
     Communication Graph is a direct graph $\mathcal{G}(V, E)$. The node set $V=\left\{ 1,...,M \right\}$ is the set of BSs, and the edge set $E$ represents the communication between BSs. There is an edge from node $i$ to node $j$ if $\text{BS}_{i}$ transmits information to $\text{BS}_{j}$. We denote the outgoing and incoming nerghbors of node $i$ by $\mathcal{N}_{O}(i)$ and $\mathcal{N}_{I}(i)$, respectively. The communication graph is said to be strongly connected if for every pair of nodes $(i,j)$ there exists a directed path from $i$ to $j$.
\end{definition}

\begin{figure}[!h]
\centering
\includegraphics[width=2.7in]{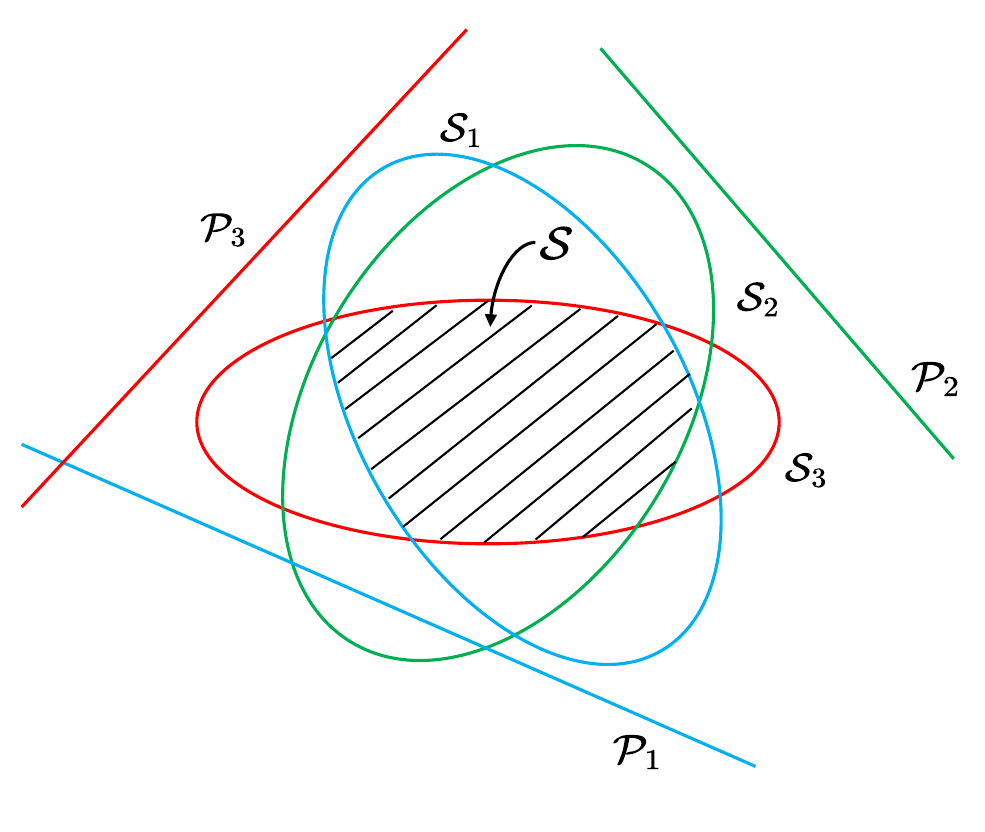}
\caption{local feasible regions $\mathcal{S}_{i}$, global feasible region $\mathcal{S}$, and cutting planes $\mathcal{D}_{i}$ that are generated by $\text{BS}_{i}$.}
\label{feasible}
\end{figure}

In \cite{6600797}, authors consider that each processor $i$ has its own constraint set and the global feasible set is the intersection of all these sets. Inspired by this design, we first distribute constraints $\| \hat{\boldsymbol{V}}_{m} \|_{F}^{2}- P_m,\forall m$ to all BSs. As for the constraint $g(\boldsymbol{V} ,\boldsymbol{\Delta} ) \leq \varepsilon$, we utilize all BSs to generate different cutting planes in order to accelerate the approximation of the feasible region. Thus, it can approximate the feasible region better and faster. Mathematically, the problem solved by $\text{BS}_{m}$ is given as
\begin{equation}
\begin{array}{cl}
    \underset{ \boldsymbol{V},  \boldsymbol{\Delta}}{\min}  &-f( \boldsymbol{V} , \boldsymbol{\Delta} ) \\
    \text { s.t. }  & \| \hat{\boldsymbol{V}}_{m}\|_{F}^{2} \leq P_{m}  \\
     &g(\boldsymbol{V} ,\boldsymbol{\Delta} ) \leq \varepsilon,
\end{array}
\end{equation}
where the feasible region is denoted by $\mathcal{S}_{m}$. Taking Figure \ref{feasible} as an example, cutting planes $\mathcal{D}_{i}, i=1,...,3$ are generated by three different BSs every $k_{pre}$ iteration in order to approximate its own feasible region $\mathcal{S}_{i}$ simultaneously and will be passed between BSs. Three cutting plane constraints combined can approximate global feasible region $\mathcal{S}$ more precisely. Compared with BLRBF which generates only one cutting plane constraint every $k_{pre}$ iteration, this distributed algorithm can generate three cutting plane constraints. The details of BLADRBF are presented in algorithm \ref{distributed_algo}. The subscript $l$ represents variables and sets at $\text{BS}_{l}$.

\begin{algorithm}[tb]
\caption{BLADRBF: BiLevel based Asynchronous Distributed Robust BeamForming.}
\label{distributed_algo}
\textbf{Input}: $\boldsymbol{P}$, $\left \{ \boldsymbol{\widetilde{h}}_{k_m n}, \epsilon_{k_m n} \right \} $ .\\
\textbf{Output}: $\boldsymbol{V}, \boldsymbol{\Delta}$.
\begin{algorithmic}[1] 
\FOR{each $\text{BS}_{l}$}
    \STATE Initialize iteration $t=0$, variables $\boldsymbol{V}_l^{[0]}, \boldsymbol{\Delta}_l^{[0]}$ with different values and dual variables $\left \{ \lambda_{l,i}^{[0]} \right \} =0$;
\ENDFOR
\FOR{each $\text{BS}_{l}$}
    \REPEAT
        \STATE  Each BS update variables $\boldsymbol{V}_l^{[t+1]}, \boldsymbol{\Delta}_l^{[t+1]}$ like algorithm \ref{centralized_algo};
        \IF {$t$ mod $k_{pre}==0$}
            \STATE It transmits its current $\mathcal{D}_{l}^{[t+1]}$ to all its out-neighbors $\mathcal{N}_{O}(l)$ and receives active constrains of its in-neighbors $Y_l^{[t+1]}=\bigcup_{j \in \mathcal{N}_{I}(l)} \mathcal{D}_j^{[t+1]}$;
        
            \STATE $\mathcal{D}_{l}^{[t+1]} \leftarrow  \mathcal{D}_{l}^{[t+1]} \bigcup Y_l^{[t+1]}$;
        \ENDIF
        \STATE $t \leftarrow t+1$;
    \UNTIL{convergence}.
\ENDFOR
\end{algorithmic}
\end{algorithm}

Note that the algorithm \ref{distributed_algo} operates without the need for time synchronization. Each BS has the flexibility to conduct its own computations at varying speeds and can update its constraints as soon as it receives relevant cutting planes. The asynchronous distributed method is more robust against communication failures.

\begin{proposition}\label{Proposition}
    The linear constraints at a BS form a polyhedral approximation of the feasible region $\mathcal{S}$.
\end{proposition}
\noindent The detailed proof of Proposition \ref{Proposition} is given in Appendix C.
\begin{theorem}\label{Theorem_distributed}
    (Consistent) Let $\overline{F}_{i}$ be the convergence value computed by $\text{BS}_{i}$. We have that $\overline{F}_{i} = \overline{F}_{j} = \overline{F}, \forall i,j$.
\end{theorem}
\noindent The proof of Theorem \ref{Theorem_distributed} is presented in Appendix D.

\section{Experiments}
In this section, numerical simulations are carried out to illustrate the performance of BLRBF and BLADRBF algorithms. We consider multi-cell multi-user MISO downlink systems. The specific parameters including the number of cells $M$, the number of users $K$, the number of antennas $N$, and the transmit power budget $\boldsymbol{P}$  are provided alongside corresponding figures. We adopt a typical small-scale fading channel model,i.e., Rayleigh fading, which is widely used in previous literature \cite{6153325,9479773}. 

\begin{figure}[!t]
\centering
\includegraphics[width=3.3in]{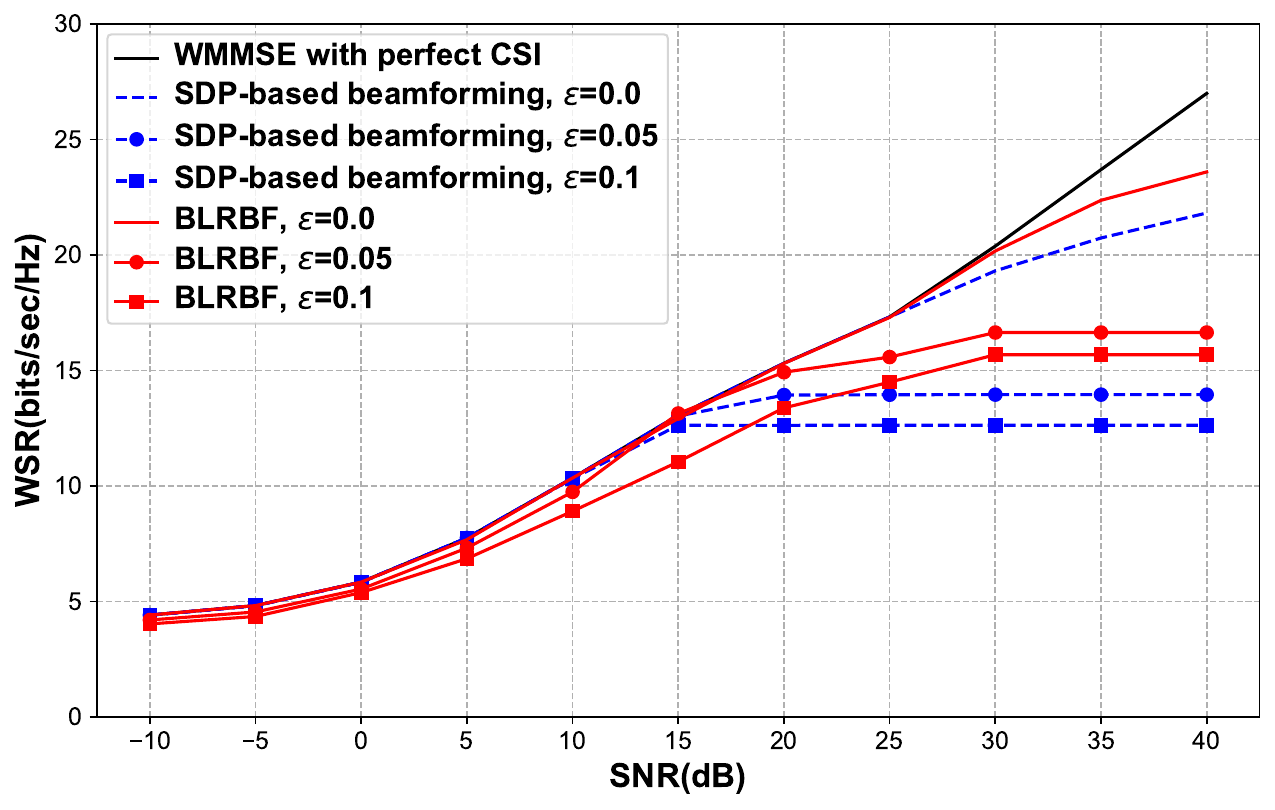}
\caption{Comparing the worst-case weighted sum-rates yielded by BLRBF, SDP-based beamforming and WMMSE algorithms for $M=N=K=2$.}
\label{fig_222}
\end{figure}

\textbf{Rayleigh fading:} Each channel coefficient $\boldsymbol{h}_{k_m n}$ is generated according to a complex standard normal distribution, i.e.,
\begin{equation*}
    Re(\boldsymbol{h}_{k_m n}) \sim \frac{\mathcal{CN}(\boldsymbol{0},\boldsymbol{I})}{\sqrt{2}}, Im(\boldsymbol{h}_{k_m n}) \sim \frac{\mathcal{CN}(\boldsymbol{0},\boldsymbol{I})}{\sqrt{2}}, \forall m,n,k.
\end{equation*}

We contrast the proposed algorithms with the SDP-based beamforming proposed in \cite{5590310}. Under the assumption of perfect CSI, the results are compared with the WMMSE method proposed in \cite{5756489}. All simulation experiments are executed on a machine equipped with a 16-core AMD Ryzen 7 5800H processor.

We perform an analysis of the robust WSR maximization across three systems with parameters $M=N=K=2$, $M=K=3, N=4$ and $M=4, K=10, N=64$. For each of these system configurations, we consider 10 erroneous channel realizations to illustrate the performance of the proposed algorithms.

Figure \ref{fig_222} displays the optimized worst-case sum-rate achieved by BLRBF, SDP-based beamforming, and WMMSE algorithms. If CSI is assumed to be perfectly known, i.e., $\epsilon_{k_m n}=0$ for all $k$, $m$, $n$, we find that both BLRBF and SDP-based beamforming algorithms can achieve comparable performance with the WMMSE algorithm, while BLRBF approach slightly outperforms the SDP-based beamforming method. Next, we explore BLRBF and SDP-based beamforming algorithms for different uncertainty regions with radii of 0.05 and 0.1. The key observation is that larger uncertainty regions lead to diminished robust weighted sum rates. This is primarily because larger regions of uncertainty indicate more significant channel estimation errors, which consequently lead to a reduced worst-case weighted sum rate. Moreover, as SNR increases, the BLRBF method outperforms the SDP-based beamforming method in terms of achieving significant improvements in robust weighted sum rates.

\begin{figure}[!t]
\centering
\includegraphics[width=3.3in]{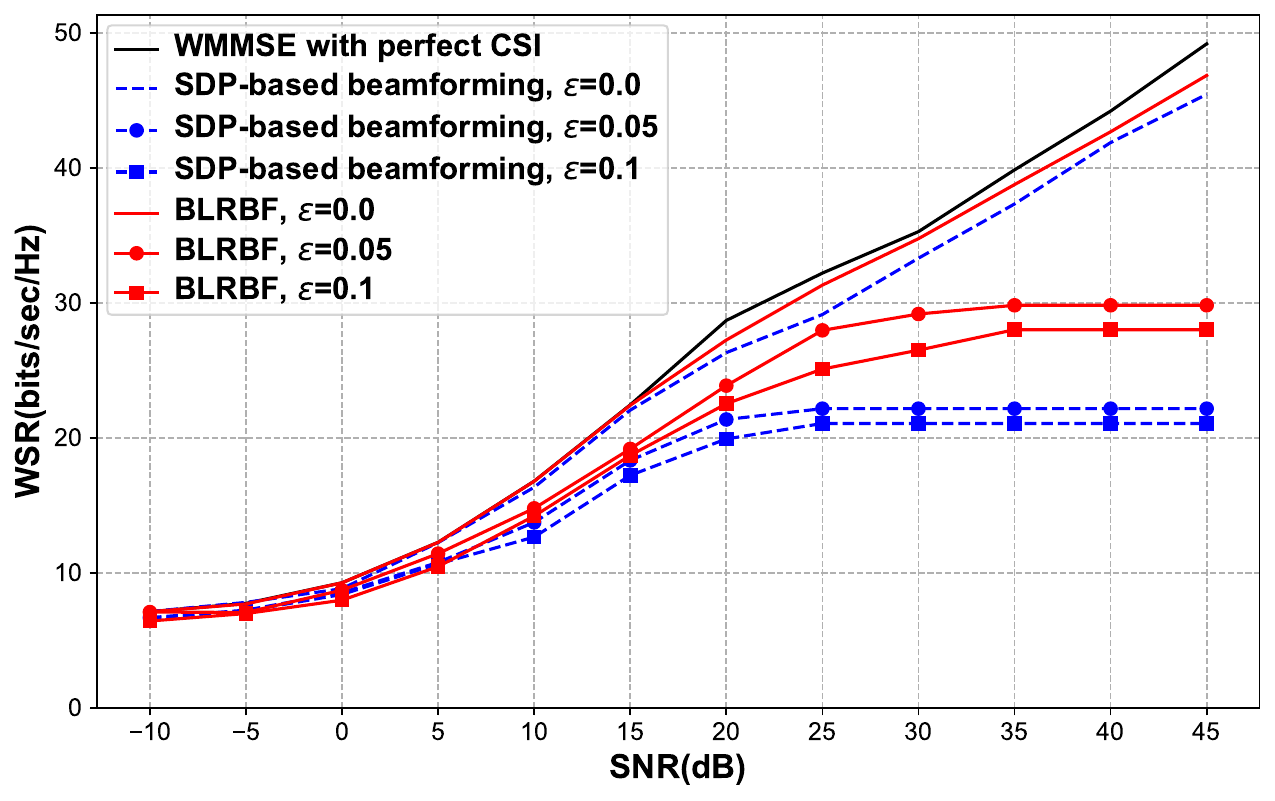}
\caption{Comparing the worst-case weighted sum-rates yielded by BLRBF, SDP-based beamforming and WMMSE algorithms for $M=K=3, N=4$.}
\label{fig_334}
\end{figure}

Continuing, we scale up the system, focusing on a configuration with $M=K=3$ and $N=4$. Figure \ref{fig_334} illustrates the relationship between the robust weighted sum rate and SNR, consistent with Figure \ref{fig_222}. This is because higher SNR means more base station power $\boldsymbol{P}$, assuming constant noise. In communication systems, greater SNR lowers noise interference, enhancing data reliability and boosting the weighted sum rate. Additionally, we observe that the optimized robust weighted sum rate saturates at high SNR levels which aligns with the high SNR analysis and Theorem 7 in \cite{5590310}. Despite both BLRBF and SDP-based beamforming algorithms saturating at high SNR for $\epsilon\neq 0$, it is noteworthy that the BLRBF method achieves saturation at a higher SNR level, and its saturation value surpasses that of SDP-based beamforming method. It is also revealed that at the same SNR, the BLRBF algorithm outperforms the SDP-based beamforming algorithm, yielding an optimized robust weighted sum rate that is superior.

\begin{figure}[!t]
\centering
\includegraphics[width=3.3in]{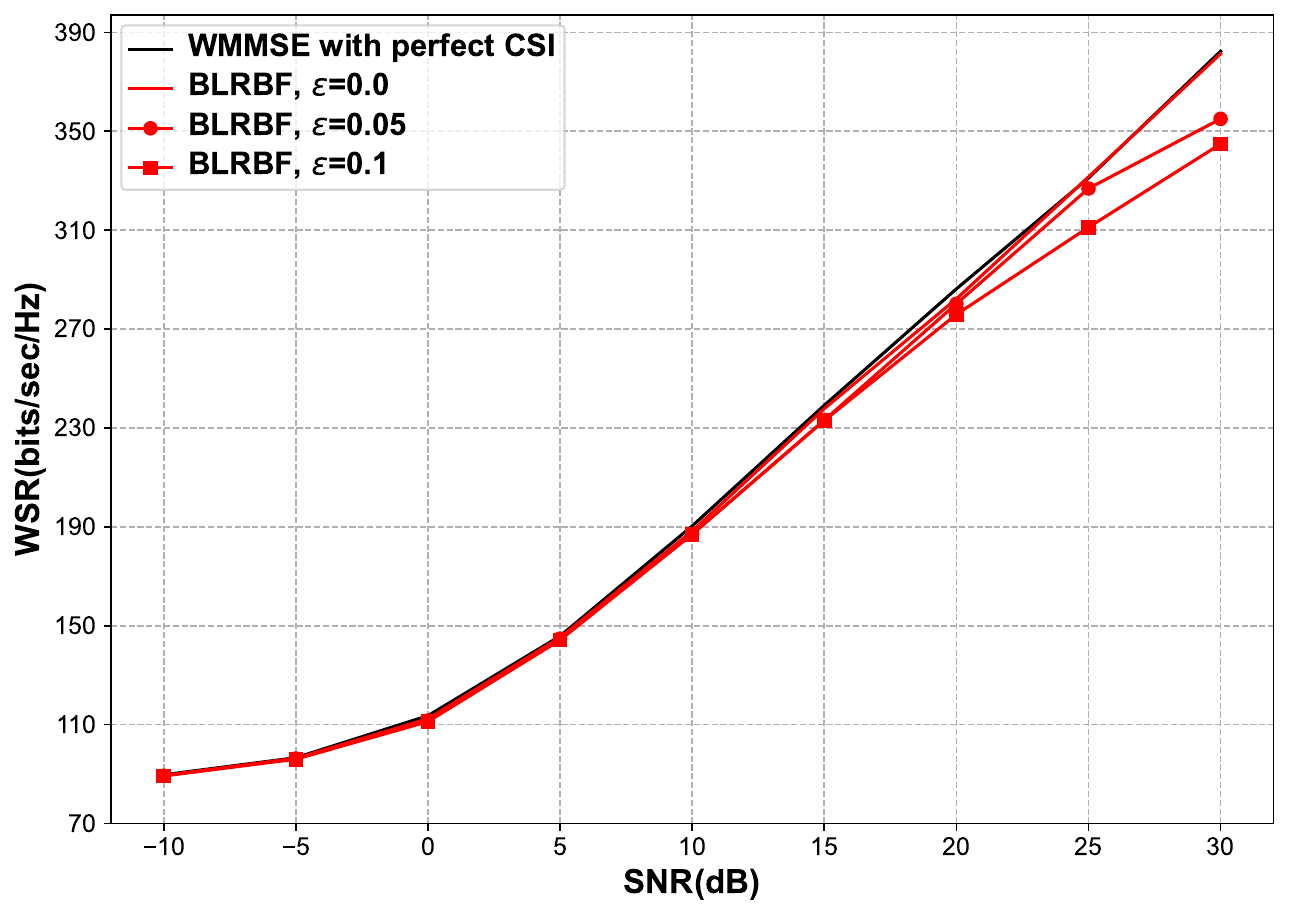}
\caption{The worst-case weighted sum-rates yielded by BLRBF algorithms for $M=4, K=10, N=64$.}
\label{fig_41064}
\end{figure}

Lastly, we delve into the analysis of a network with a more extensive setup, that is $M=4, K=10, N=64$. This setup emulates a more realistic large-scale scenario. When the matrix size is large, SDP solved through the interior point method becomes computationally expensive and even intolerable. Specifically, the computation time of the SDP-based beamforming method exceeds $4$ hours under the scenario settings of $M=4, K=10, N=64$. In contrast, our proposed BLRBF approach can efficiently achieve the optimized results within an acceptable time frame. In Figure \ref{fig_41064}, the worst-case weighted sum rates versus SNRs for this setup are depicted. This analysis demonstrates the capability of our proposed BLRBF algorithm to handle large-scale scenarios effectively, even in situations where the SDP-based beamforming method faces computational challenges.

\begin{figure}[!t]
\centering
\includegraphics[width=3.3in]{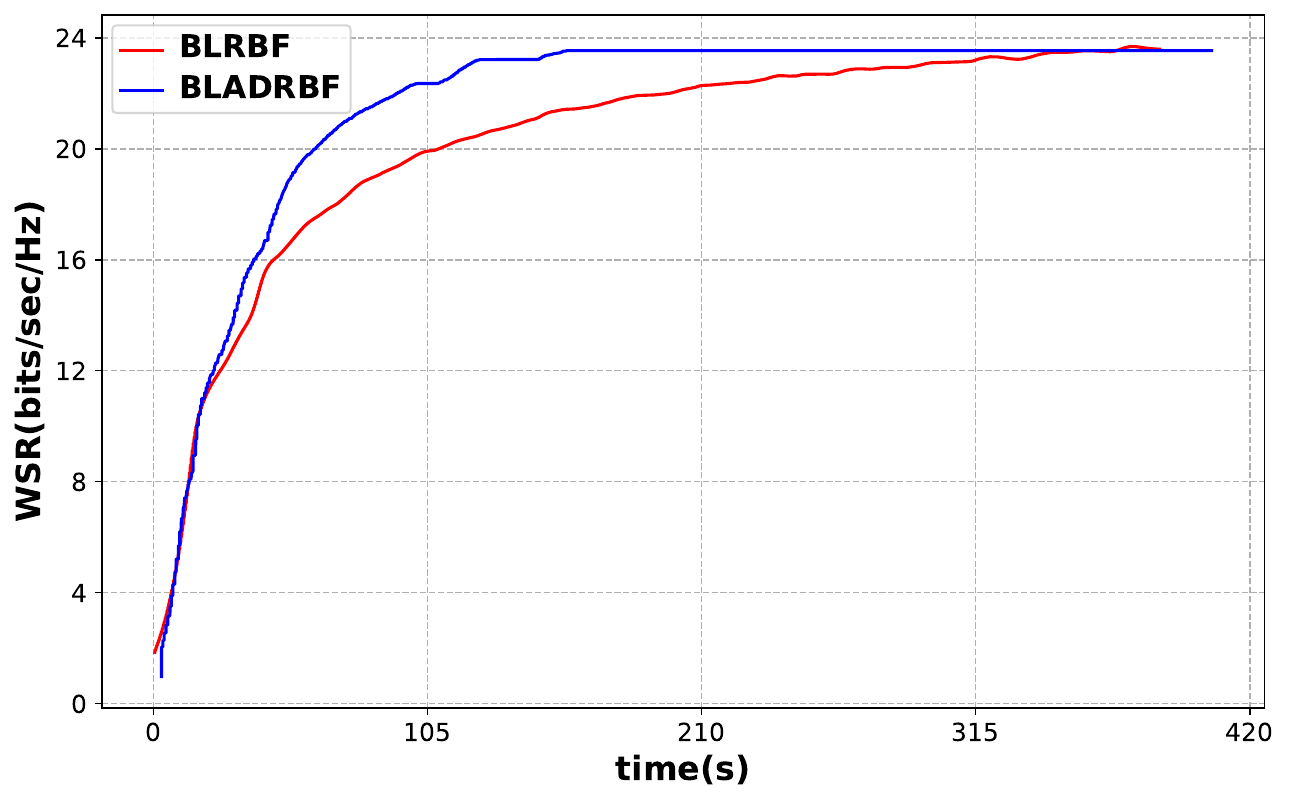}
\caption{Comparing the convergence rate of BLRBF and BLADRBF algorithms for $M=3, K=3, N=4$.}
\label{fig_dis}
\end{figure}

Finally, we demonstrate the disparity in terms of convergence rates between the BLRBF algorithm and the BLADRBF algorithm in Figure \ref{fig_dis}. It is evident that while both BLRBF and BLADRBF algorithms attain similar final convergence values, the BLADRBF approach exhibits a more rapid convergence.

\section{Conclusion}
We propose to address robust beamforming problems by adopting a bilevel optimization perspective, thereby providing a fresh insight into this field. Focusing on the problem of maximizing the worst-case weighted sum-rate for multi-cell multi-user MISO wireless networks where BSs can acquire only noisy channel estimates, we develop an efficient algorithm, i.e., BLRBF, based on the cutting plane method. A distributed algorithm called BLADRBF is also proposed to facilitate the parallel processing in practical settings. We prove both algorithms are guaranteed to converge. Our algorithm can be readily extended to MIMO systems. Finally, through comprehensive numerical experiments, we demonstrate that the BLRBF method can significantly outperform the SDP-based beamforming method proposed in \cite{5590310}, particularly in high SNR regimes. We also confirm that the distributed algorithm BLADRBF exhibits a faster convergence rate compared to the centralized algorithm BLRBF.

\section{Acknowledgments}
This work was supported in part by the National Natural Science Foundation of China under
Grant 12371519 and 61771013; in part by the Fundamental Research Funds for the Central Universities of China; and in part by the Fundamental Research Funds of Shanghai Jiading District.

\bibliography{aaai24}

\begin{thebibliography}{35}
\providecommand{\natexlab}[1]{#1}

\bibitem[{Björnson, Bengtsson, and Ottersten(2014)}]{6832894}
Björnson, E.; Bengtsson, M.; and Ottersten, B. 2014.
\newblock Optimal Multiuser Transmit Beamforming: A Difficult Problem with a Simple Solution Structure [Lecture Notes].
\newblock \emph{IEEE Signal Processing Magazine}, 31(4): 142--148.

\bibitem[{Bogale and Vandendorpe(2012)}]{6104172}
Bogale, T.~E.; and Vandendorpe, L. 2012.
\newblock Weighted Sum Rate Optimization for Downlink Multiuser MIMO Coordinated Base Station Systems: Centralized and Distributed Algorithms.
\newblock \emph{IEEE Transactions on Signal Processing}, 60(4): 1876--1889.

\bibitem[{Boyd and Vandenberghe(2007)}]{boyd2007localization}
Boyd, S.; and Vandenberghe, L. 2007.
\newblock Localization and cutting-plane methods.
\newblock \emph{From Stanford EE 364b lecture notes}, 386.

\bibitem[{Bürger, Notarstefano, and Allgöwer(2014)}]{6600797}
Bürger, M.; Notarstefano, G.; and Allgöwer, F. 2014.
\newblock A Polyhedral Approximation Framework for Convex and Robust Distributed Optimization.
\newblock \emph{IEEE Transactions on Automatic Control}, 59(2): 384--395.

\bibitem[{Choi et~al.(2012)Choi, Park, Lee, and Lee}]{6153325}
Choi, H.-J.; Park, S.-H.; Lee, S.-R.; and Lee, I. 2012.
\newblock Distributed Beamforming Techniques for Weighted Sum-Rate Maximization in MISO Interfering Broadcast Channels.
\newblock \emph{IEEE Transactions on Wireless Communications}, 11(4): 1314--1320.

\bibitem[{Franceschi et~al.(2018)Franceschi, Frasconi, Salzo, Grazzi, and Pontil}]{franceschi2018bilevel}
Franceschi, L.; Frasconi, P.; Salzo, S.; Grazzi, R.; and Pontil, M. 2018.
\newblock Bilevel programming for hyperparameter optimization and meta-learning.
\newblock In \emph{International conference on machine learning}, 1568--1577. PMLR.

\bibitem[{Ji et~al.(2020)Ji, Lee, Liang, and Poor}]{ji2020convergence}
Ji, K.; Lee, J.~D.; Liang, Y.; and Poor, H.~V. 2020.
\newblock Convergence of Meta-Learning with Task-Specific Adaptation over Partial Parameters.
\newblock In \emph{Advances in Neural Information Processing Systems}, 11490--11500.

\bibitem[{Ji, Yang, and Liang(2021)}]{ji2021bilevel}
Ji, K.; Yang, J.; and Liang, Y. 2021.
\newblock Bilevel optimization: Convergence analysis and enhanced design.
\newblock In \emph{International conference on machine learning}, 4882--4892. PMLR.

\bibitem[{Jiao, Yang, and Song(2022)}]{NEURIPS2022_34899013}
Jiao, Y.; Yang, K.; and Song, D. 2022.
\newblock Distributed Distributionally Robust Optimization with Non-Convex Objectives.
\newblock In \emph{Advances in Neural Information Processing Systems}, 7987--7999.

\bibitem[{Jiao et~al.(2023)Jiao, Yang, Wu, Song, and Jian}]{jiao2022asynchronous}
Jiao, Y.; Yang, K.; Wu, T.; Song, D.; and Jian, C. 2023.
\newblock Asynchronous distributed bilevel optimization.
\newblock In \emph{International Conference on Learning Representations}.

\bibitem[{Jorge and Stephen(2006)}]{jorge2006numerical}
Jorge, N.; and Stephen, J.~W. 2006.
\newblock \emph{Numerical optimization}.
\newblock Spinger.

\bibitem[{Joshi et~al.(2012)Joshi, Weeraddana, Codreanu, and Latva-aho}]{6119233}
Joshi, S.~K.; Weeraddana, P.~C.; Codreanu, M.; and Latva-aho, M. 2012.
\newblock Weighted Sum-Rate Maximization for MISO Downlink Cellular Networks via Branch and Bound.
\newblock \emph{IEEE Transactions on Signal Processing}, 60(4): 2090--2095.

\bibitem[{Joudeh and Clerckx(2016)}]{7555358}
Joudeh, H.; and Clerckx, B. 2016.
\newblock Sum-Rate Maximization for Linearly Precoded Downlink Multiuser MISO Systems With Partial CSIT: A Rate-Splitting Approach.
\newblock \emph{IEEE Transactions on Communications}, 64(11): 4847--4861.

\bibitem[{Liu, Simonyan, and Yang(2018)}]{liu2018darts}
Liu, H.; Simonyan, K.; and Yang, Y. 2018.
\newblock Darts: Differentiable architecture search.
\newblock In \emph{International Conference on Learning Representations}.

\bibitem[{Liu, Zhang, and Chua(2012)}]{6170850}
Liu, L.; Zhang, R.; and Chua, K.-C. 2012.
\newblock Achieving Global Optimality for Weighted Sum-Rate Maximization in the K-User Gaussian Interference Channel with Multiple Antennas.
\newblock \emph{IEEE Transactions on Wireless Communications}, 11(5): 1933--1945.

\bibitem[{Liu et~al.(2021)Liu, Liu, Yuan, Zeng, and Zhang}]{liu2021value}
Liu, R.; Liu, X.; Yuan, X.; Zeng, S.; and Zhang, J. 2021.
\newblock A value-function-based interior-point method for non-convex bi-level optimization.
\newblock In \emph{International Conference on Machine Learning}, 6882--6892. PMLR.

\bibitem[{Luo and Zhang(2008)}]{4453890}
Luo, Z.-Q.; and Zhang, S. 2008.
\newblock Dynamic Spectrum Management: Complexity and Duality.
\newblock \emph{IEEE Journal of Selected Topics in Signal Processing}, 2(1): 57--73.

\bibitem[{Rong, Vorobyov, and Gershman(2006)}]{1665009}
Rong, Y.; Vorobyov, S.; and Gershman, A. 2006.
\newblock Robust linear receivers for multiaccess space-time block-coded MIMO systems: a probabilistically constrained approach.
\newblock \emph{IEEE Journal on Selected Areas in Communications}, 24(8): 1560--1570.

\bibitem[{Shaverdian and Nakhai(2014)}]{6824195}
Shaverdian, A.; and Nakhai, M.~R. 2014.
\newblock Robust Distributed Beamforming With Interference Coordination in Downlink Cellular Networks.
\newblock \emph{IEEE Transactions on Communications}, 62(7): 2411--2421.

\bibitem[{Shen et~al.(2012)Shen, Chang, Wang, Qiu, and Chi}]{6156468}
Shen, C.; Chang, T.-H.; Wang, K.-Y.; Qiu, Z.; and Chi, C.-Y. 2012.
\newblock Distributed Robust Multicell Coordinated Beamforming With Imperfect CSI: An ADMM Approach.
\newblock \emph{IEEE Transactions on Signal Processing}, 60(6): 2988--3003.

\bibitem[{Shenouda and Davidson(2008)}]{4586299}
Shenouda, M.~B.; and Davidson, T.~N. 2008.
\newblock On the Design of Linear Transceivers for Multiuser Systems with Channel Uncertainty.
\newblock \emph{IEEE Journal on Selected Areas in Communications}, 26(6): 1015--1024.

\bibitem[{Shi et~al.(2011)Shi, Razaviyayn, Luo, and He}]{5756489}
Shi, Q.; Razaviyayn, M.; Luo, Z.-Q.; and He, C. 2011.
\newblock An Iteratively Weighted MMSE Approach to Distributed Sum-Utility Maximization for a MIMO Interfering Broadcast Channel.
\newblock \emph{IEEE Transactions on Signal Processing}, 59(9): 4331--4340.

\bibitem[{Sun et~al.(2022)Sun, Pu, Fu, Chang, and Hong}]{9682542}
Sun, H.; Pu, W.; Fu, X.; Chang, T.-H.; and Hong, M. 2022.
\newblock Learning to Continuously Optimize Wireless Resource in a Dynamic Environment: A Bilevel Optimization Perspective.
\newblock \emph{IEEE Transactions on Signal Processing}, 70: 1900--1917.

\bibitem[{Tajer, Prasad, and Wang(2011)}]{5590310}
Tajer, A.; Prasad, N.; and Wang, X. 2011.
\newblock Robust Linear Precoder Design for Multi-Cell Downlink Transmission.
\newblock \emph{IEEE Transactions on Signal Processing}, 59(1): 235--251.

\bibitem[{Tran et~al.(2012)Tran, Hanif, Tolli, and Juntti}]{6327333}
Tran, L.-N.; Hanif, M.~F.; Tolli, A.; and Juntti, M. 2012.
\newblock Fast Converging Algorithm for Weighted Sum Rate Maximization in Multicell MISO Downlink.
\newblock \emph{IEEE Signal Processing Letters}, 19(12): 872--875.

\bibitem[{Von~Stackelberg(1934)}]{hicks1935marktform}
Von~Stackelberg, H. 1934.
\newblock Marktform und gleichgewicht.

\bibitem[{Vucic and Boche(2009)}]{4663943}
Vucic, N.; and Boche, H. 2009.
\newblock Robust QoS-Constrained Optimization of Downlink Multiuser MISO Systems.
\newblock \emph{IEEE Transactions on Signal Processing}, 57(2): 714--725.

\bibitem[{Weber, Sklavos, and Meurer(2006)}]{1605472}
Weber, T.; Sklavos, A.; and Meurer, M. 2006.
\newblock Imperfect channel-state information in MIMO transmission.
\newblock \emph{IEEE Transactions on Communications}, 54(3): 543--552.

\bibitem[{Weeraddana et~al.(2013)Weeraddana, Codreanu, Latva-aho, and Ephremides}]{6331563}
Weeraddana, P.~C.; Codreanu, M.; Latva-aho, M.; and Ephremides, A. 2013.
\newblock Multicell MISO Downlink Weighted Sum-Rate Maximization: A Distributed Approach.
\newblock \emph{IEEE Transactions on Signal Processing}, 61(3): 556--570.

\bibitem[{Xue et~al.(2021)Xue, Wang, Yan, Hu, Yang, and Sun}]{xue2021rethinking}
Xue, C.; Wang, X.; Yan, J.; Hu, Y.; Yang, X.; and Sun, K. 2021.
\newblock Rethinking bi-level optimization in neural architecture search: A gibbs sampling perspective.
\newblock In \emph{Proceedings of the AAAI Conference on Artificial Intelligence}, volume~35, 10551--10559.

\bibitem[{Yang et~al.(2014)Yang, Huang, Wu, Wang, and Chiang}]{yang2014distributed}
Yang, K.; Huang, J.; Wu, Y.; Wang, X.; and Chiang, M. 2014.
\newblock Distributed robust optimization ({DRO}), part {I}: {Framework} and example.
\newblock \emph{Optimization and Engineering}, 15(1): 35--67.

\bibitem[{Yang et~al.(2008)Yang, Wu, Huang, Wang, and Verdu}]{4509766}
Yang, K.; Wu, Y.; Huang, J.; Wang, X.; and Verdu, S. 2008.
\newblock Distributed Robust Optimization for Communication Networks.
\newblock In \emph{IEEE INFOCOM 2008 - The 27th Conference on Computer Communications}, 1157--1165.

\bibitem[{Zhang et~al.(2022)Zhang, Yuan, Zheng, Krikidis, and Wong}]{9479773}
Zhang, J.; Yuan, Y.; Zheng, G.; Krikidis, I.; and Wong, K.-K. 2022.
\newblock Embedding Model-Based Fast Meta Learning for Downlink Beamforming Adaptation.
\newblock \emph{IEEE Transactions on Wireless Communications}, 21(1): 149--162.

\bibitem[{Zhang, Palomar, and Ottersten(2008)}]{4567684}
Zhang, X.; Palomar, D.~P.; and Ottersten, B. 2008.
\newblock Statistically Robust Design of Linear MIMO Transceivers.
\newblock \emph{IEEE Transactions on Signal Processing}, 56(8): 3678--3689.

\bibitem[{Zhou et~al.(2020)Zhou, Pan, Ren, Wang, Renzo, and Nallanathan}]{9110587}
Zhou, G.; Pan, C.; Ren, H.; Wang, K.; Renzo, M.~D.; and Nallanathan, A. 2020.
\newblock Robust Beamforming Design for Intelligent Reflecting Surface Aided MISO Communication Systems.
\newblock \emph{IEEE Wireless Communications Letters}, 9(10): 1658--1662.

\end{thebibliography}

\end{document}